# MULTICHANNEL KONDO IMPURITIES IN SUPERCONDUCTORS

L.S. Borkowski and P.J. Hirschfeld

*Department of Physics, University of Florida, Gainesville, FL 32611*

We discuss the effect of multichannel Kondo impurities on superconductivity. In the strong coupling regime such impurities are pairbreakers, in contrast to the ordinary Kondo effect. Measurements of $T_c$-suppression may help in identifying impurities displaying this more exotic exchange coupling to the conduction band.

The multichannel Kondo problem is known to have a non-Fermi liquid fixed point with divergent $T \to 0$ susceptibilities. Its experimental realization in some heavy-fermion systems was proposed but remains controversial.[1–3] Here we would like to offer another test of the multichannel Kondo effect. The breaking of time-reversal symmetry involved in scattering by magnetic impurities in superconductors has important consequences. The qualitative type of behavior is well-known in the single channel problem to depend on the ratio $T_K/T_{c0}$, where $T_K$ is the Kondo temperature and $T_{c0}$ is the transition temperature of a pure superconductor. The divergence of magnetic correlation length and the existence of a residual magnetic moment as $T \to 0$ suggest that in the multichannel case one should expect a very different type of interplay between exchange and pairing interactions.

One may use a sensitivity of superconducting correlations to multichannel exchange interaction as an additional criterion to characterize multichannel Kondo behavior and distinguish it from the single channel case. In this paper we study a simplified SU(N)×SU(k) version of the full multichannel problem in the NCA approximation. Here, $N$ is the orbital degeneracy of the impurity and k is the number of conduction electron channels coupled to the impurity. Although this approach ignores details of anisotropic exchange it remains in the same universality class. In particular the exponents for the temperature dependence of the susceptibilities agree with those obtained from the conformal field theory.[3,4]

Our model includes a BCS pairing of electrons in the conduction band. In general the pairing may be either of channel singlet or triplet type. We assume the latter possibility. The quantities we study in this work are not significantly different for the channel singlet state. In the limit of large on-site Coulomb repulsion $U$ and for temperatures $T \ll U$, the model has a simplified form,

$$H = \sum_{k,\alpha,\sigma} \epsilon_k c^\dagger_{k\alpha\sigma} c_{k\alpha\sigma} + E_f \sum_\sigma f^\dagger_\sigma f_\sigma$$
$$+ V \sum_{k,\alpha,\sigma} \left[ c^\dagger_{k,\alpha,\sigma} f_\sigma b^\dagger_{\bar\alpha} + h.c. \right]$$
$$+ \sum_{k,\alpha,\sigma} \left[ \Delta(k) c^\dagger_{k,\alpha,\sigma} c^\dagger_{-k,\alpha,-\sigma} + h.c. \right]$$
$$+ \lambda \left( \sum_\sigma f^\dagger_\sigma f_\sigma + \sum_\alpha b^\dagger_{\bar\alpha} b_{\bar\alpha} - 1 \right). \qquad (1)$$

The indices $\alpha$ and $\sigma$ refer to channel and spin,

respectively. In the limit $\lambda \to \infty$ the unphysical states with the impurity occupation $n_f > 1$ are projected out. The position of the bare impurity level is assumed to be $E_f = -0.67D$, and $\Gamma = \pi N(0)V^2 = 0.15D$, where $D$ is half of the band width. For this parameter set the Kondo temperature is $T_K = 4.5 \times 10^{-5} D$ and the impurity occupation number is $n_f \simeq 0.9$.

The pairing correlations in the conduction band lead to a nonzero anomalous impurity propagator, $F_{f,\sigma\sigma'}^{\alpha\alpha'}(\tau) = \langle T_\tau f_\sigma(\tau) b_\alpha(\tau) f_{\sigma'}(0) b_{\alpha'}(0) \rangle$ (see Figure 1).

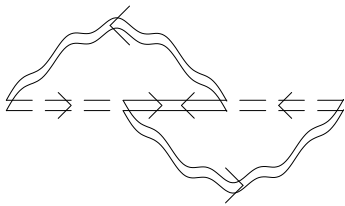

Fig. 1. The leading order diagram for the anomalous impurity propagator. The dashed lines are the fermion propagators and the wavy and solid lines are boson and conduction electrons respectively.

This propagator, including the internal conduction electron line is evaluated self-consistently at finite impurity concentrations. Although $F_f \sim 1/N^2$, and in NCA only the contributions $O(1/N)$ are retained, $F_f$ is of order $O(\Delta)$ near $T_c$ and must be included in the calculation of the superconducting transition temperature. There are no anomalous contributions to boson and fermion self-energies. We calculate $F_f$ in the "elastic" approximation[5] in which the internal anomalous conduction electron propagator is evaluated at the external frequency. The inclusion of inelastic processes is not expected to significantly affect the results because the spin-flip and non-spin-flip contributions almost cancel.[6] At low temperatures these off-diagonal components of the T-matrix have a peak at $\omega \simeq T$ (see Figure 2).

To evaluate $T_c$-suppression we use the gap

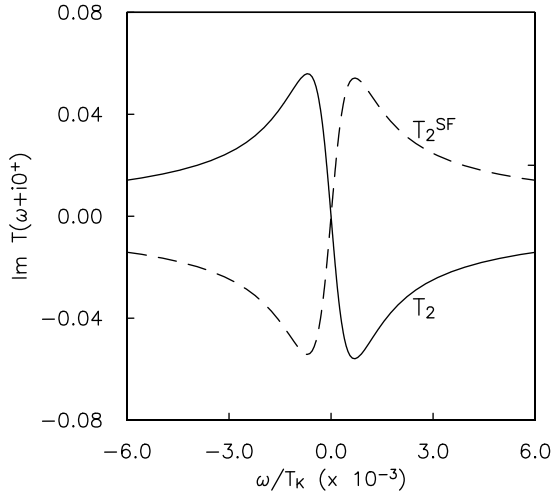

Fig. 2. The anomalous part of the conduction electron T-matrix evaluated at $T = 4.5 \times 10^{-4} T_K$. The solid (broken) line is the non-spin-flip (spin-flip) contribution.

equation,

$$1 = V_S \int_0^{\omega_D} d\omega N_0(\omega) \frac{\tilde{\Delta}}{\Delta \tilde{\omega}} \tanh(\omega/2T_c), \qquad (2)$$

where $N_0(\omega)$ is the density of conduction electron states in the normal metal and $V_S$ is the pairing potential. The full conduction electron Green's function, averaged over impurity positions, $G^{-1}(\omega) = \tilde{\omega}\tau_0 - \epsilon_k \tau_3 - \tilde{\Delta}(k)\tau_1$, is found from the Dyson equations, $\tilde{\omega} = \omega - nV^2 G_f(\tilde{\omega})$, and $\tilde{\Delta} = \Delta + nV^2 F_f(\tilde{\omega})$, where $n$ is impurity concentration.

As shown in Fig. 3, the slope $N_0(0) dT_c/dn$, at $n = 0$, remains finite as $T_K/T_{c0} \to \infty$, indicating finite pairbreaking in the strong coupling regime. The qualitative form of the $T_c$ dependence on $n$ remains practically unchanged for all values of $T_K/T_{c0}$, however, as illustrated in Fig. 4.

Both results are a consequence of finite magnetic moment of the impurity at all temperatures. The maximum slope is reached for $T_K/T_{c0} \to \infty$, in contrast with the single channel Kondo problem where maximum occurs for $T_K \simeq T_{c0}$. Similar results for $dT_c/dn$ were obtained in a re-cent Monte Carlo calculation.[6] If the Kondo scale is known from other experiments, measurements of $T_c$ as a function of impurity concentration may help in identifying the multichannel behavior. For $T_{c0} \lesssim T_K$, studies of the single channel problem have found a reentrant or exponential dependence of $T_c$ on concentration.[7] If it were possible to vary the ratio $T_K/T_{c0}$ in the strong coupling regime, increased pairbreaking with growing $T_K/T_{c0}$ would imply multichannel behavior while the opposite would be true for a single channel coupling. In unconventional superconductors, for which $\sum_k F(k,\omega) = 0$, both single and multichannel exchange result in qualitatively similar form of the $T_c$-dependence on $n$ in the strong coupling regime and the differences which may exist for $T_K \simeq T_{c0}$ may be harder to indentify. Finally, let us note that the breaking of the spin or channel symmetry which takes the system away from the multichannel fixed point should reduce pairbreaking especially in the strong coupling limit. Given the subtleties of this analysis we conclude that $T_c$-suppression is not likely to be a sensitive tool for identifying the multichannel exchange.

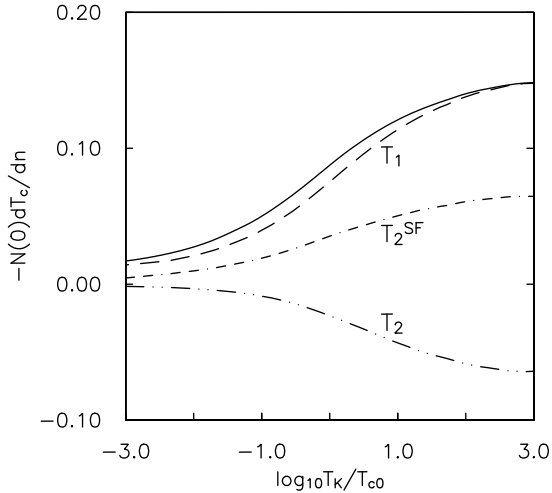

Fig. 3. The slope of the initial $T_c$-suppression with separate contributions from the diagonal part of the T-matrix, $T_1$, the off-diagonal spin-flip, $T_2^{SF}$, and the off-diagonal non-spin-flip, $T_2$. The solid line is the sum of the three contributions.

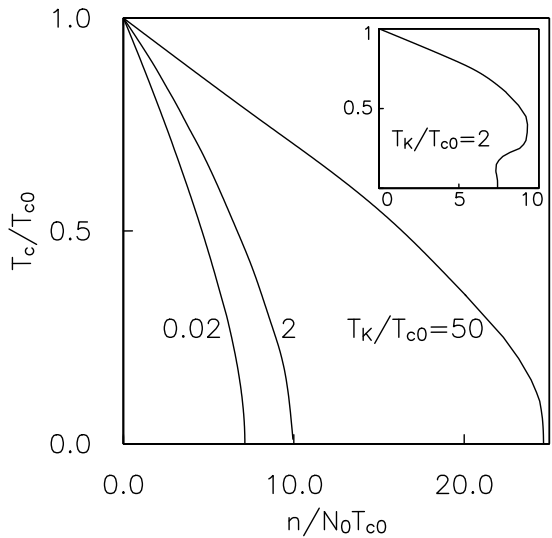

Fig. 4. The superconducting transition temperature as a function of impurity concentration. The inset shows the result for a single channel Kondo impurity with $T_K \simeq T_{c0}$ evaluated with $E_f = -0.67D$, and $\Gamma = 0.15D$.

We are grateful to K. Ingersent for discussions and H. Kroha and M. Hettler for advice on numerical calculations. We would like to thank D. Cox for bringing Ref. [6] to our attention.